\RequirePackage{fix-cm}
\documentclass[twocolumn]{svjour3}          
\smartqed  
\usepackage{cite}

\usepackage[english]{babel}

\usepackage{graphicx}

\usepackage{grffile}
\usepackage{epstopdf}

\usepackage[tight,footnotesize]{subfigure}

\usepackage{fixltx2e}

\usepackage[cmex10]{amsmath}
\usepackage{color}
\usepackage{array}
\usepackage{enumerate}
\usepackage{amssymb}
\begin{document}

\title{Interference Calculation in Asynchronous Random Access Protocols using Diversity
}


\author{Alessio Meloni         \and
        Maurizio Murroni 
}

\authorrunning{A. Meloni et al.} 

\institute{Department of Electrical and Electronic Engineering (DIEE), University of Cagliari \at
              Piazza D'Armi, 09123 Cagliari, Italy \\
              \email{\{alessio.meloni\}\{murroni\}@diee.unica.it}  \\         
This paper has been accepted for publication in the Springer's Telecommunication Systems journal. The final publication will be made available at Springer. Please refer to that version when citing this paper. DOI: 10.1007/s11235-015-9970-3
}

\date{Received: 12 Sept. 2014 / Accepted: 17 Jan. 2015}

\maketitle

\begin{abstract}
The use of Aloha-based Random Access protocols is interesting when channel sensing is either not possible or not convenient and the traffic from terminals is unpredictable and sporadic. In this paper an analytic model for packet interference calculation in asynchronous Random Access protocols using diversity is presented. The aim is to provide a tool that avoids time-consuming simulations to evaluate packet loss and throughput in case decodability is still possible when a certain interference threshold is not exceeded. Moreover the same model represents the groundbase for further studies in which iterative Interference Cancellation is applied to received frames. 
\end{abstract}

\section{Introduction} \label{Intro}

Random Access protocols present great advantages in multiaccess scenarios when the traffic is sporadic and/or hardly predictable \cite{yue} \cite{tian}. As a matter of fact, they avoid signalling overhead and long transmission latencies as in reservation based mechanisms. For this reason Aloha-based schemes have attracted interest in many different fields such as car-to-car, mobile and satellite communications \cite{CRDSA2} \cite{stabMEL2}.
 
In the original idea \cite{AbramsonALOHA} terminals send packets as soon as they are received from the upper layers, without the need of coordination among them. This introduces the possibility of collision among bursts sent from different terminals but ensures an acceptable Packet Loss Ratio when the traffic from terminals is low while minimizes both packet latency and system complexity. The best average throughput achievable in this case is $T\simeq0.18$. 

Its slotted evolution known as Slotted Aloha (SA) \cite{RobertsALOHA} introduces synchronization among terminals so that the channel is divided into slots and terminals aiming to transmit data start their transmission in a synchronous manner. As a result, the throughput is doubled ($T\simeq0.36$) at the cost of slightly increased packet delay and bigger complexity of the system. This is obtained thanks to synchronization that eliminates the case of partial interference thus minimizing the overall number of collisions. 

A further extension of SA known as Diversity Slotted Aloha (DSA) \cite{DiversityALOHA} introduces the concept of packet replication as a mean to create time diversity. While this increases the physical load on the channel and thus the probability of collision of the single copy, the probability to receive at least one of the redundant copies without collision is augmented for small channel loads.   

Recently, the use of Interference Cancellation (IC) in Aloha-based techniques has been introduced \cite{CRDSA1} as a mean to further exploit diversity. This new technique, called Contention Resolution Diversity Slotted Aloha (CRDSA) consists in clearing slots from the content of the remaining copies of already decoded bursts thanks to their location's knowledge provided by pointers in the decoded packet. As a result, not only bursts received without interference are decoded but also those colliding with a burst that has a decodable twin, thus boosting the throughput up to $T\simeq0.97$ depending on the number of copies sent per packet and on the channel load.

Successively the same concept has been extended to the case of asynchronous terminals \cite{CRA} \cite{ECRA} demonstrating that the joint use of time diversity and IC yields to results that are comparable and, depending on the SNR, even better than those obtained with synchronous RA techniques \cite{stabMEL3}.

In \cite{IRSA1} an analytic model for CRDSA was also presented in order to compute packet loss and throughput based on the probability that a certain packet's copy was interfering or not. In the case of asynchronous RA schemes such a model has not been developed yet, even though it would be useful to study the related stability as in \cite{kumar} and \cite{stabMEL1}. Moreover, in the asynchronous case the concept of interference is broaden since not only complete but also partial interference is possible \cite{lau2010}. 

This work focuses on interference calculation for asynchronous Aloha-based RA protocols using diversity. A similar work was done in \cite{zorzi} for the case of Slotted Aloha in which the only possibility is that packets are entirely interfering. The aim of this paper is to provide a tool for packet loss and throughput evaluation for the case in which decodability is still possible when a certain interference threshold is not exceeded. This represents an important step forward for the design and dimensioning of the system, usually requiring time consuming simulations and lacking an analytic framework that can allow the designer to set constraints and maximize the performance of the channel analytically based on the formulas. Moreover the same model represents the groundbase for further studies in which iterative IC can be applied to received frames. 

The remainder of this paper is organized as follows: in section II an overview of the considered system is given and the problem that arises from the lack of an analytic framework for interference calculation is stated; Section III defines the decoding threshold used to declare a burst still decodable; Section IV introduces the mathematical model for interference calculation; in section V numerical results aiming at validating the described model are presented; Section VI concludes the paper and discusses future work in light of the obtained results. 

\section{System Overview and Problem Statement}\label{PS}

\begin{figure}[h!]
\centering
\includegraphics [width=1 \columnwidth] {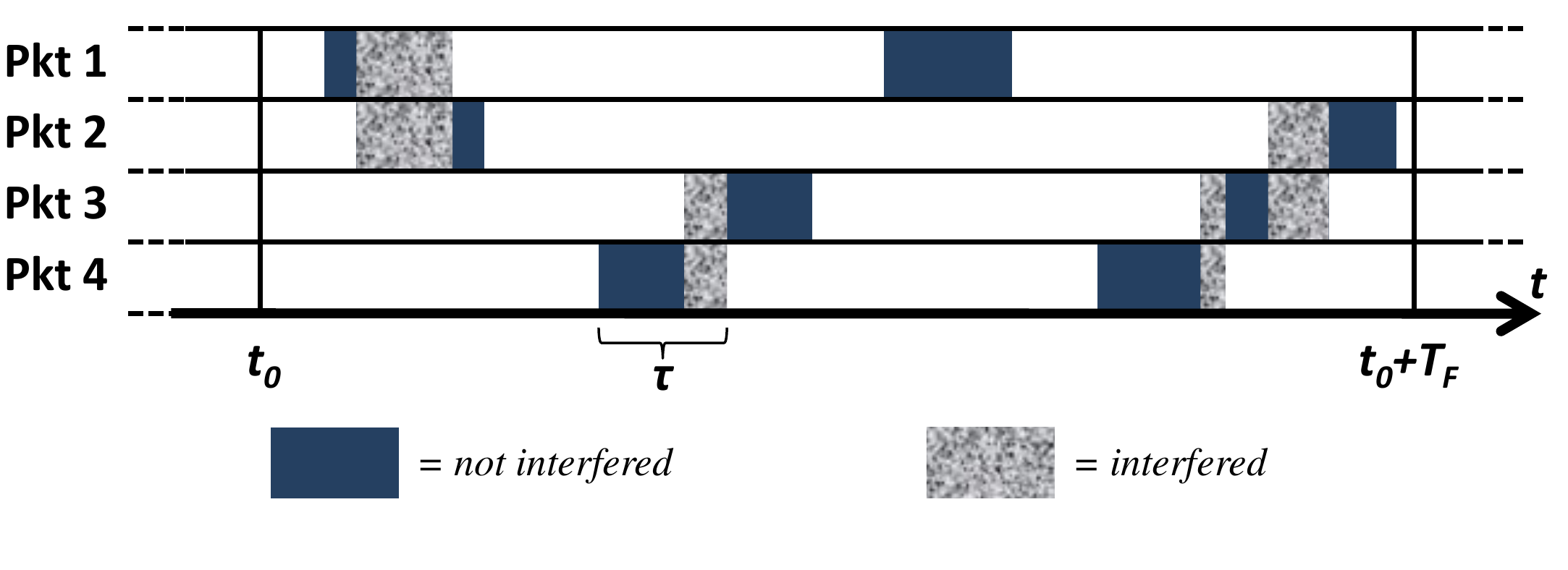}
\caption{Example of a generic frame at the receiver.}
\label{CRA}
\end{figure}

Consider a SC-TDMA channel divided into frames. When a terminal has a packet to transmit, it waits for the beginning of the next available frame in order to place $d$ non-overlapping copies of its packet. Let us call $t_0$ the beginning of a frame, $T_F$ its time duration and $\tau$ the timelength of a burst\footnote{In this work all burst durations are assumed to be equal.}. The $d$ copies of a packet are placed with starting time uniformly distributed within the interval [$t_0$ , $t_0+T_F-\tau$]. In the example in Figure~\ref{CRA} four packets are sent in the same frame and two copies are placed for each packet.

When the frame arrives at the receiver, each copy can have a certain degree of interference due to transmission time overlap with bursts from the other terminals. Assuming equal power among received bursts, if no FEC is used any interference will result in loss of the involved burst. However, if a strong FEC is used decodability will still be possible as long as the amount of interference does not exceed a certain threshold. 

For example, in Figure~\ref{CRA} $Pkt\ 1$ has a copy that did not interfere with other bursts. Therefore assuming no noise or other sources of disturbance, it will be correctly received. However if we consider the application of a strong FEC with a certain interference threshold for decodability (e.g. no more than $\tau/3$ overlapping among two bursts), we can see that also the content of $Pkt\ 3$ and $Pkt\ 4$ will be correctly received.
 
The amount of packets sent in a frame is measured as the normalized MAC channel load\footnote{With the term normalized MAC channel load we refer to the load generated from different packets regardless of the number of copies generated for each packet.}, defined as
\begin{equation}
G=\frac{N_{tx}\cdot \tau}{T_F}
\end{equation}
with $N_{tx}$ indicating the number of transmitted packets. The amount of data packets correctly received at the destination is measured as 
\begin{equation}
T(G)=G [1-PLR(G)]
\end{equation}
which represents the throughput in terms of portion of load successfully decoded. The term PLR represents the Packet Loss Ratio which is usually computed through simulations rather than analytically. For this reason in the next sections, after defining the decoding threshold, a model for interference calculation that allows analytical computation of the PLR will be presented.

\section{Decoding Threshold Definition}
Apart from the load $G$, the PLR depends on the frame size $T_F$, the burst size $\tau$, the number of copies per packet $d$ and the decoding threshold that, for the sake of comparison, has been defined as done in \cite{CRA} and to which we refer the reader for the discussion on this choice.
Consider the rate $R$ to be
\begin{equation}
R=R_C\cdot log_2M
\end{equation}
 where $R_C$ is the coding rate and $M$ the modulation index. Approximating the decoding threshold with the Shannon bound\footnote{As claimed in \cite{CRA}, even though this threshold is quite optimistic it can be considered valid for moderate to high SNIR. Nevertheless the developed model can be easily adapted by simply substituting a different value for the threshold.} we can set the channel capacity to
\begin{equation} 
C=R=log_2(1+SNIR)
\end{equation}
so that the decoding threshold is 
\begin{equation}
SNIR_{dec,dB}=10\cdot log_{10}(2^R-1)
\end{equation}

In order for a burst to be decoded, its $SNIR$ must be at least equal to $SNIR_{dec}$. The $SNIR$ of each burst can be computed as

\begin{equation}
SNIR=\frac{P}{(x/\tau)\cdot P+N}=\frac{SNR}{(x/\tau)\cdot SNR+1}
\end{equation}
where $x$ represents the degree of interference. For example, in case of no interference $x=0$, in case of total overlapping with another burst $x=\tau$ and in the more general case of partial interference with $n$ other bursts, $x=x_1+x_2+\cdot\cdot\cdot+x_n$ where $x_1$, $x_2$, $\cdot\cdot\cdot$, $x_n$ represent the degree of interference due to each overlapping burst, expressed with a value between $0$ and $\tau$. 

\section{Interference Model}

First of all we want to know the amount of interference between the generic copy of a packet called hereinafter Considered Copy (CC) and the $d$ copies of another packet, that we call Disturbing Packet (DP). In the followings we assume symbol synchronization among terminals and the symbol time $T_s$ is considered to be our discrete time unit, so that $x$, $\tau$ and $T_F$ can be expressed as multiples of $T_s$:
\begin{equation}
x=x'\cdot T_s
\end{equation}
\begin{equation}
\tau=\tau'\cdot T_s
\end{equation}
\begin{equation}
T_F=T_F'\cdot T_s
\end{equation}
with $\{x', \tau', T_F'\} \in\mathbb{N}$\\

The CC can be in one of three states:
\begin{itemize}
\item the whole CC interfers with one or more DP's copies, i.e. $x'=\tau'$;
\item partial interference of the CC with one or more DP's copies, i.e. $x'\in\{1,2,...,\tau'-1\}$;
\item no interference with any of the considered DP's copies, i.e. $x'=0$.\\
\end{itemize}

Moreover, each single DP's copy can generate 4 different events named (A),(B),(C) and (D):

\begin{enumerate}[(A)]
\item the DP's copy completely overlaps with the CC;
\item the DP's copy partially overlaps with the CC;
\item the DP's copy does not overlap with the CC and the number of possible interference outcomes for the other DP's copies remains unmodified;
\item the DP's copy does not overlap with the CC but number of possible interference outcomes for the other DP's copies is modified.
\end{enumerate}

\begin{figure}[h!]
\centering
\includegraphics [width=1 \columnwidth] {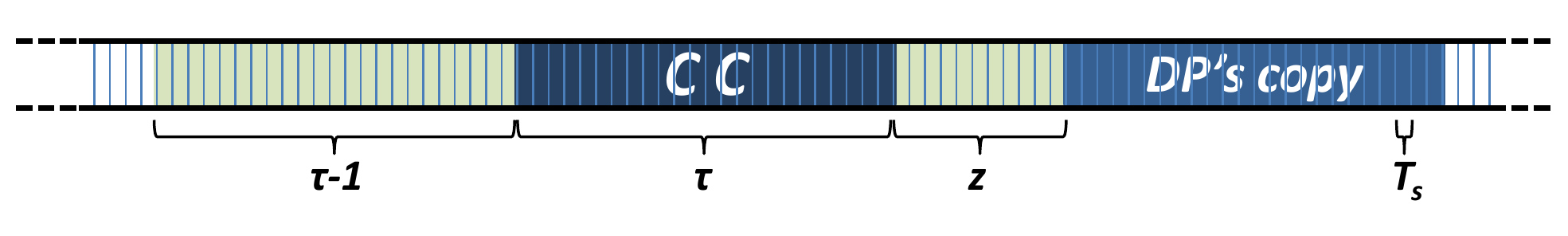}
\caption{Example of non-overlapping that modifies the interference outcomes for the other DP's copies.}
\label{CD}
\end{figure}

The reason for the distinction between the events (C) and (D) is due to the fact that even though no interference occurs between the CC and the considered DP's copy, if the two bursts are distant less than $\tau$ then some of the interference outcomes for the remaining copies are not possible anymore (see Figure~\ref{CD}). The distance between the CC and a certain DP's copy can be defined as $z$ that once again is a multiple of the symbol time, i.e.

\begin{equation}
z=z'\cdot T_s
\end{equation}

\subsection{Interference contribution of one DP}

In \cite{CRA} it was shown that in the case of good $SNR$ ($10 dB$) the highest throughput for CRA is achieved for $d=2$. Moreover, even though the case of $d=2$ is not the best one for low $SNR$ ($2dB$), the following two considerations can be made. 1) The best curve in the case of low $SNR$ ($d=3$) and in general all the curves with $d>2$, degrade more rapidly than the one for $d=2$ after the peak value. This represents an hazard in case of statistical fluctuations of the load that will tend to drive the channel to saturation. 2) In some application scenarios such as WSN or multi-access channels via bent-pipe satellite, limitations on the average power impose that the same budget for information packet is used thus resulting in poor performance when the $SNR$ is low and the number of copies per packet is greater than $2$ \cite{normeff}. In light of these considerations, in this work we decided to concentrate on the calculation for the case of $d=2$.

Let us indicate $X$ and $Y$ as two generic events generated by the DP's copies. In the followings, the caption $(X,Y)_{x'}$ will be used to indicate the space of the events in which the event generated by the first copy of the DP is $X$, the event generated by the second copy of the DP is $Y$ and the overall resulting interference on the CC is $x'$. In other words,
\begin{equation} \begin{split}
(X,Y)_{x'}=\sum_\delta(X)_{\delta}\cup (Y|X)_{x'-\delta}
\end{split} \end{equation}\\
 Therefore, as an example, all the cases in which the two DP's copies have partial interference with the CC and the overall effect is a CC completely interfered are indicated as $(B,B)_{\tau'}$. Finally, considering other two generic events W and Z, the captions $(XW,Y)_{x'}$ and $(X,YZ)_{x'}$  are used to indicate $(X,Y)_{x'}\cup(W,Y)_{x'}$ and  $(X,Y)_{x'}\cup(X,Z)_{x'}$ respectively. 
Given this nomenclature, the probability of the event $(X,Y)_{x'}$ can be expressed as $Pr\{(X,Y)_{x'}\}$. 

In the followings we consider the case in which the CC is not close to the edges of the frame. As a matter of fact, if the CC is closer than $\tau$ to the end or the start of the frame, the number of possible interference outcomes is reduced thus making our analytical description a lower bound. Although it is possible to take into account this case, this would further complicate the model. Moreover for realistic cases $T_F\gg \tau$, so that the effect due to the edges of the frame results negligible. This will be shown in the results section.\\

\subsubsection{Total Interference}
The probability that the CC has completely interfered with one or more copies of the DP is\\
\begin{equation} \begin{split}
Pr\{x'=\tau'\}=Pr\{(A,CD)_{\tau'}\}+\\ \\
+Pr\{(CD,A)_{\tau'}\}+Pr\{(B,B)_{\tau'}\}
\end{split} \end{equation}\\

 where\\
\begin{equation}\label{eqA1}
Pr\{(A,CD)_{\tau'}\}=\frac{1}{T_F'-\tau'+1}
\end{equation}\\
is the probability that the first DP's copy was completely interfering and thus the second copy could not interfere at all,
\begin{equation}\label{eqA2}
Pr\{(CD,A)_{\tau'}\}=\frac{1}{T_F'-\tau'+1}
\end{equation}\\
is the probability that the first DP's copy did not interfere at all while the second copy was completely interfering.

Equation \ref{eqA1} comes from\\

\begin{equation} \begin{split}
Pr\{(A,CD)_{\tau'}\}=Pr\{(A)_{\tau'}\}\cdot Pr\{(CD|A)_{0}\}= \\ \\
=\frac{1}{T_F'-\tau'+1} \cdot \frac{T_F'-\tau'+1-(2\tau'-1)}{T_F'-\tau'+1-(2\tau'-1)}
\end{split} \end{equation}\\

and similarly equation \ref{eqA2} comes from\\

\begin{equation} \begin{split}
Pr\{(CD,A)_{\tau'}\}=Pr\{(CD|A)_{0}\}\cdot Pr\{(A)_{\tau'}\}=\\ \\
=\frac{T_F'-\tau'+1-(2\tau'-1)}{T_F'-\tau'+1} \cdot \frac{1}{T_F'-\tau'+1-(2\tau'-1)}
\end{split} \end{equation}\\

while the last term is

\begin{equation}\label{eq13} \begin{split}
Pr\{(B,B)_{\tau'}\}=\sum_{\delta=1}^{\tau'-1}Pr\{(B)_{\delta}\}\cdot Pr\{(B|B)_{\tau'-\delta}\}=\\ \\
=\frac{2(\tau'-1)}{(T_F'-\tau'+1)}\cdot \frac{1}{(T_F'-\tau'+1-(2\tau'-1))}
\end{split} \end{equation} \\

Equation \ref{eq13} comes from the fact that there are $2(\tau'-1)$ occurrences in which the first DP's copy can be placed causing partial interference and $1$ occurence in which the second copy can complement interference with the whole CC.\\

\subsubsection{Partial Interference}
The probability that the CC is partially interfering with one or more copies of the DP is\\
\begin{equation} \begin{split}
Pr\{x'\}=Pr\{(B,CD)_{x'}\}+Pr\{(B,B)_{x'}\}+\\ \\
+Pr\{(C,B)_{x'}\}+Pr\{(D,B)_{x'}\} 
\end{split} \end{equation}\\ \\ for $x'=1,2,...,\tau'-1$\\

where each term is thoroughly explained in the followings.\\

\begin{equation} \begin{split}
Pr\{(B,CD)_{x'}\}=\frac{2}{T_F'-\tau'-1}\cdot\\ \\
\cdot \frac{T_F'-\tau'+1-(2\tau'-1)-(\tau'-x')}{T_F'-\tau'+1-(2\tau'-1)}
\end{split} \end{equation}\\
In Equation (19) the first copy have 2 possible occurrences of partial interference $x'$ while the second DP's copy has $T_F'-\tau'+1-(2\tau'-1)-(\tau'-x')$ no interference occurrences.\\ 
\begin{equation} \begin{split}
Pr\{(B,B)_{x'}\}=2\cdot (x'-1)\cdot \frac{1}{T_F'-\tau'+1}\cdot\\ \\
\cdot \frac{1}{T_F'-\tau'+1-(2\tau'-1)}
\end{split} \end{equation}\\ \\
In Equation (20) it is shown that partial interference $x'$ can be the overall result of $2$ partially interfering copies. This can take place in $2(x'-1)$ ways. This formula can be considered as a generalization of Equation \ref{eq13}.\\
\begin{equation} \begin{split}
 Pr\{(C,B)_{x'}\}=\frac{T_F'-\tau'+1-(4\tau'-1)}{T_F'-\tau'+1}\cdot\\ \\
\cdot \frac{2}{T_F'-\tau'+1-(2\tau'-1)}
\end{split} \end{equation}\\
In Equation (21) the first factor represents the union of all the events (C) for the first copy while the second factor represents $Pr\{(B|C)_{x'}\}$.\\
\begin{equation} \begin{split}
Pr\{(D,B)_{x'}\}=\frac{2x'}{T_F'-\tau'+1}\cdot \frac{2}{T_F'-\tau'+1-(2\tau'-1)} +\\ \\
+ \frac{2(\tau'-x')}{T_F'-\tau'+1}\cdot \frac{1}{T_F'-\tau'+1-(2\tau'-1)}\\
\end{split} \end{equation}\\
Equation (22) means that given the event $(D)$ for the first copy, there are $2x'$ cases in which the second replica has two possible occurrences of partial interference $x'$ and $2(\tau'-x')$ cases in which the second replica has one possible occurrence of partial interference $x'$.\\
\subsubsection{No Interference}
The probability that the CC does not interfere with any of the DP's copies is\\
\begin{equation} \begin{split}
Pr\{x'=0\}=Pr\{(C,CD)_0\}+Pr\{(D,CD)_0\}
\end{split} \end{equation}\\
where the two terms are explained below.\\
\begin{equation} \begin{split}
Pr\{(C,CD)_0\}=\frac{T_F'-\tau'+1-(4\tau'-1)}{T_F'-\tau'+1}\cdot \\ \\
\cdot \frac{T_F'-\tau'+1-2(2\tau'-1)}{T_F'-\tau'+1-(2\tau'-1)}
\end{split} \end{equation}\\
In Equation (24), the first factor represents the union of all the events (C) while the second factor represents all the events without interference for the second copy.
\begin{equation} \begin{split}
Pr\{(D,CD)_0\}=\sum_{z'=0}^{\tau'-1} \frac{2}{T_F'-\tau'+1}\cdot \\ \\
\cdot \frac{T_F'-\tau'+1-(3\tau'+z'-1)}{T_F'-\tau'+1-(2\tau'-1)}
\end{split} \end{equation}\\
The equation above considers the case in which, despite no interference, an additional number of occurrences are not possible due to the proximity of the CC with the first DP's copy.\\

\subsection{Interference contribution of more than one DP}

Since each terminal places its copies independently from the others, from the point of view of the CC any DP's interference distribution is represented by the same probability mass function. Therefore the resulting interference distribution due to 2 DPs can be calculated as the autoconvolution of $Pr\{x'\}$, the interference distribution of 3 DPs can be calculated as the convolution of $Pr\{x'\}$ with the interference distribution of 2 DPs and so on. In general, let us call $P_{int}^{(N)}$ the interference distribution probability mass function due to \textit{N} DPs and composed of $(N\cdot\tau'+1)$ elements. The resulting probability mass function can be defined recursively as

\begin{equation} \begin{split}
P_{int}^{(N)}=P_{int}^{(N-1)}* P_{int}^{(1)}
\end{split} \end{equation}

\subsection{PLR derivation}

In Section III the decoding threshold was defined. Based on Equations (5) and (6) it is possible to calculate $x_{dec}$ that is the interference threshold for decodability. Therefore, once $P_{int}^{(N_{tx}-1)}$ has been calculated, the probability that the CC will be correctly decoded can be computed as

\begin{equation} \begin{split}
P_{ccd}=\sum_{i=0}^{x_{dec}} P_{int}^{(N_{tx}-1)}(i)
\end{split} \end{equation}

and the probability that at least one of the $2$ copies of the same packet will be correctly decoded is

\begin{equation} \begin{split}
P_{pd}=1-(1-P_{ccd})^2
\end{split} \end{equation}

\section{Numerical Results}
In this section, the results obtained through the analytic model are presented and compared with simulations.
The parameter values initially chosen for simulations are $T_s=1\mu s$, $T_F=10^5\mu s$, $\tau=10^3\mu s$, $M=4$, $R_C=1/2$, and a number of simulation rounds per channel load point equal to $10^4$.

\begin{figure}[htbp!]
\centering
\includegraphics [width=1 \columnwidth] {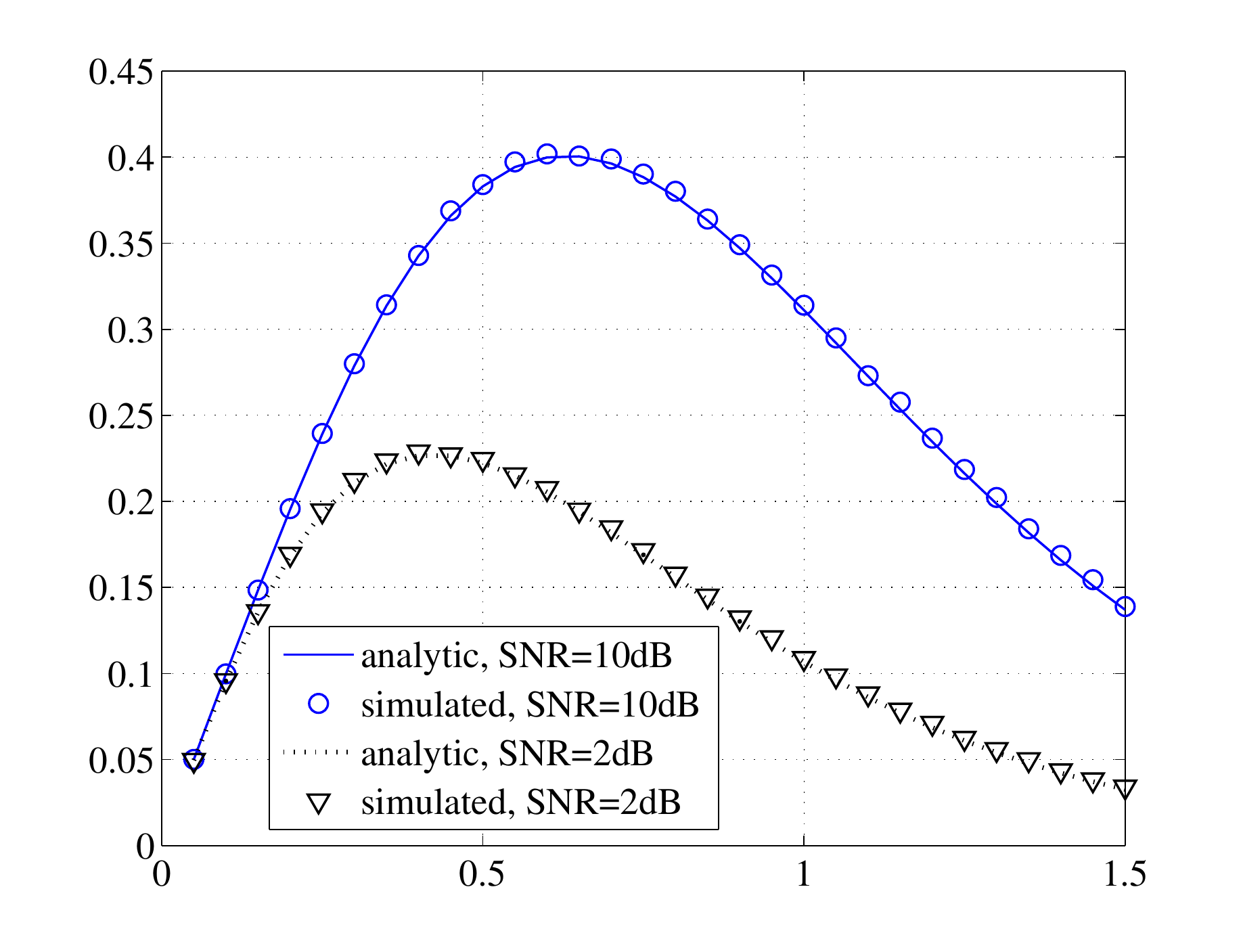}
\caption{Throughput for $T_s=1\mu s$, $T_F=10^5\mu s$, $\tau=10^3\mu s$, $M=4$, $R_C=1/2$.}
\label{THRP100}
\end{figure}

\begin{figure}[htbp!]
\centering
\includegraphics [width=1 \columnwidth] {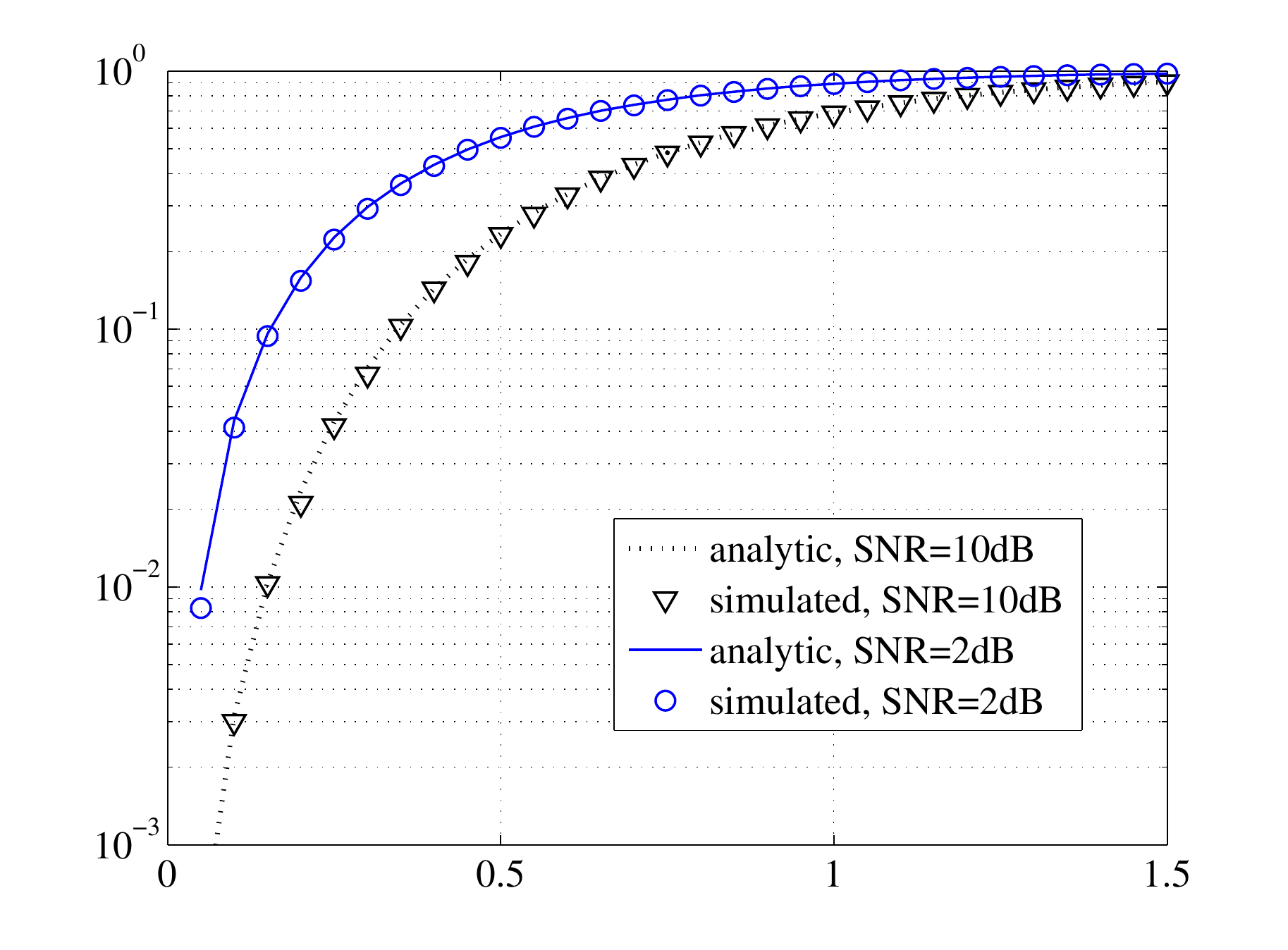}
\centering
\caption{Packet Loss Ratio for $T_s=1\mu s$, $T_F=10^5\mu s$, $\tau=10^3\mu s$, $M=4$, $R_C=1/2$.}
\label{PLR100}
\end{figure}

Figures \ref{THRP100} and \ref{PLR100} show respectively throughput and PLR results both for the analytic model and for simulations. As we can see, regardless of the $SNR$ the analytical curves obtained tightly match those gathered from simulation, proving that the interference model presented in the previous section tightly describes the ongoing inteference patterns.

\begin{figure}[htbp!]
\centering
\includegraphics [width=1 \columnwidth] {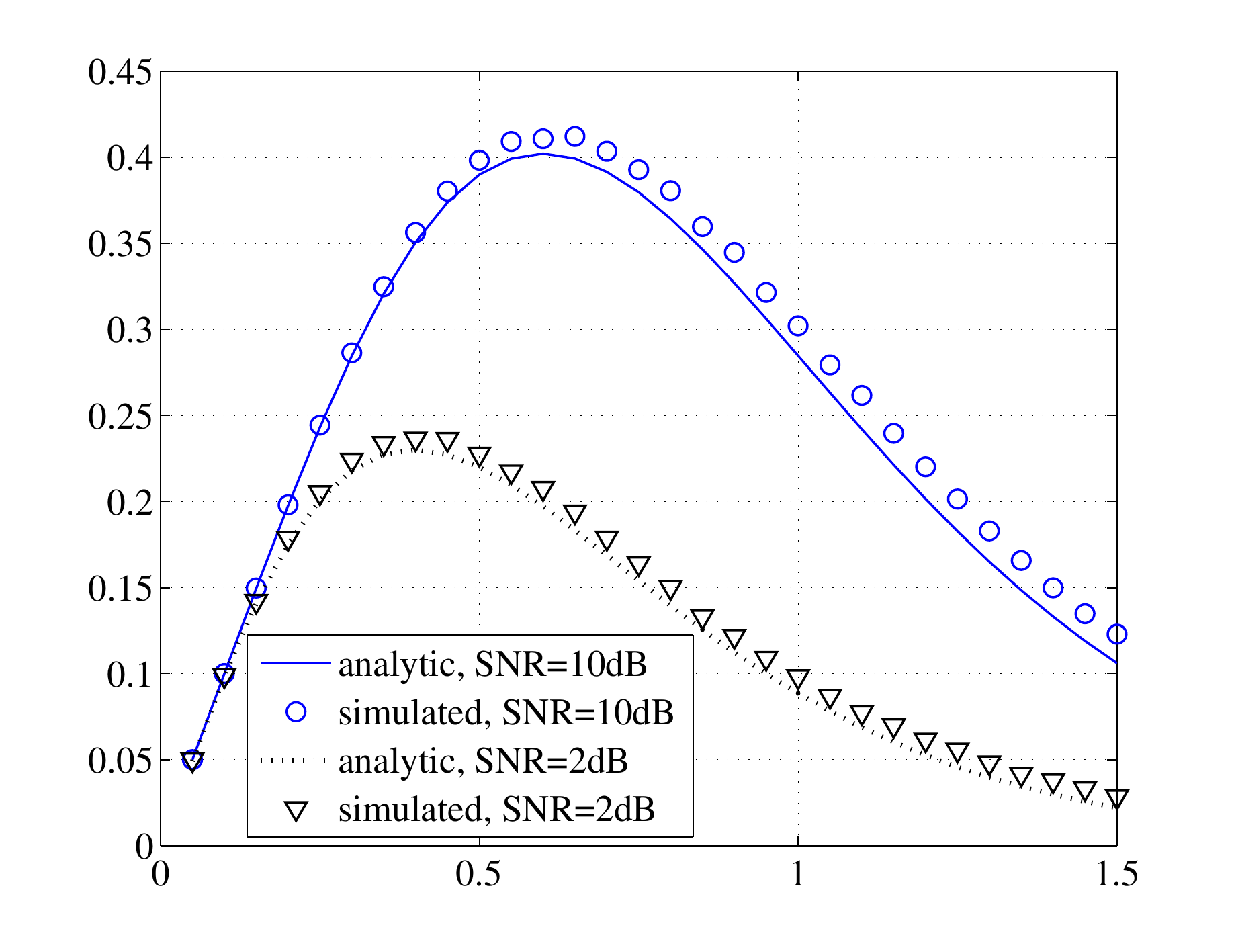}
\caption{Throughput for $T_s=1\mu s$, $T_F=2\cdot10^4\mu s$, $\tau=10^3\mu s$, $M=4$, $R_C=1/2$.}
\label{THRP20}
\end{figure}

\begin{figure}[htbp!]
\centering
\includegraphics [width=1 \columnwidth] {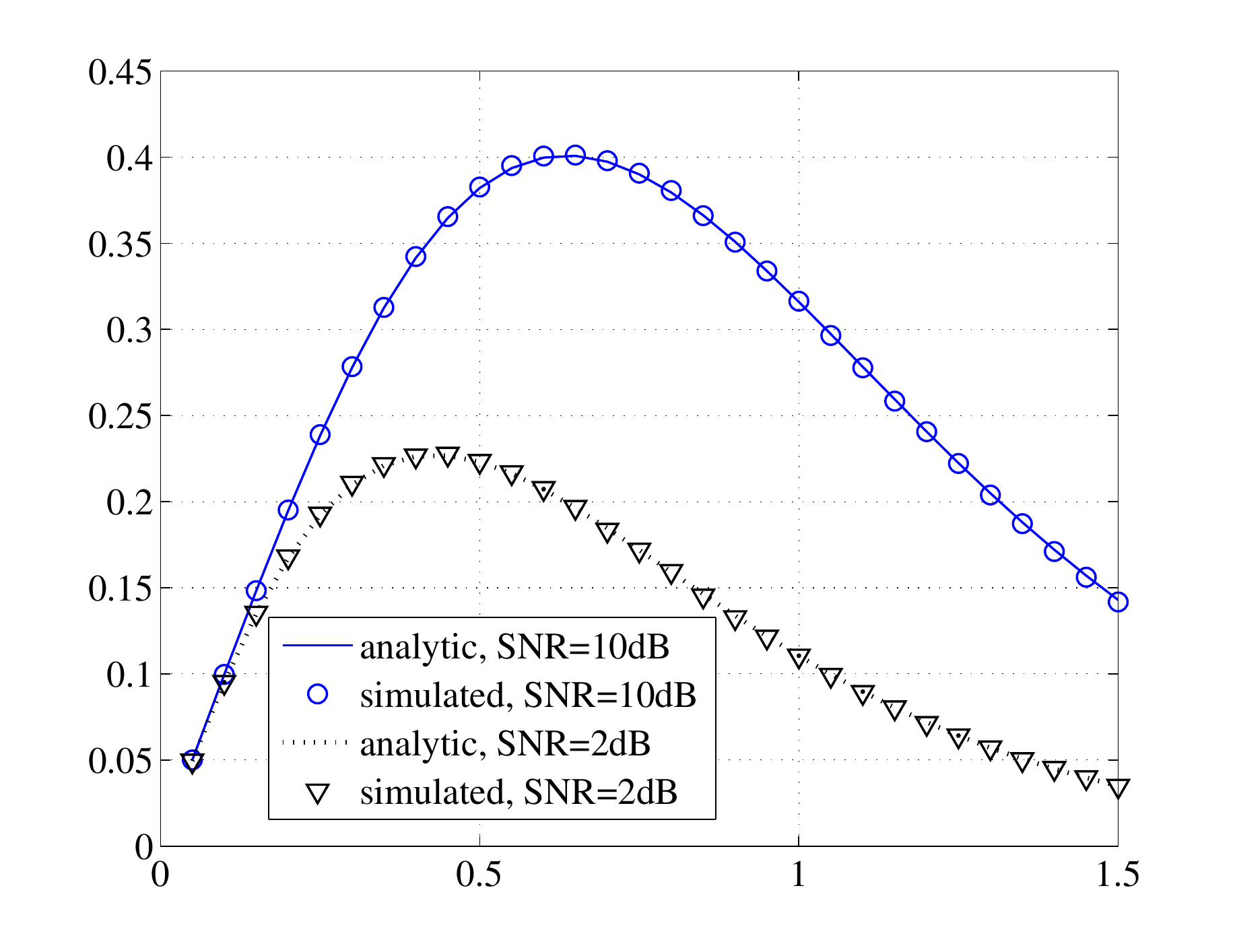}
\centering
\caption{Throughput for $T_s=1\mu s$, $T_F=2\cdot10^5\mu s$, $\tau=5\cdot10^2\mu s$, $M=4$, $R_C=1/2$.}
\label{THRP200}
\end{figure}

Concerning the approximation due to the edges of the frame that has been outlined in the previous section, figure \ref{THRP20} shows how this approximation stops being negligible when the condition $T_F\gg \tau$ is not anymore satisfied. As we can see, in this case the analytic curve becomes a lower bound rather than a tight description, in light of the fact that when the bursts are placed close to the edges of the frame their interference probability is reduced. Therefore the greater is the ratio between $T_F$ and $\tau$, the more tha analytic description is tight to the simulated results as shown in Figure \ref{THRP200} for $T_F=2\cdot10^5$ and $\tau=5\cdot10^2$.

\section{Conclusions and future work}
Unslotted protocols offer several advantages over slotted ones such as packet length flexibility and loose terminal synchronization constraints. Recently, following similar progress in the context of synchronous Aloha-based techniques, the use of interference cancellation and the use of a strong FEC have been introduced in asynchronous Aloha-based techniques that use packet diversity as a mean to exploit partial interference recovering bursts that do not exceed a certain interference threshold. In this paper an analytic model for interference calculation in the abovementioned scenario has been introduced. This model allows to calculate packet loss and throughput without the need of time consuming simulations, thus representing an important step forward towards the design and dimensioning of the system. Numerical results have shown that the presented model tightly describes simulated results under the constraint that the frame size is much greater than the burst size. Nevertheless, this condition generally holds in real communication scenarios using Aloha-like transmission schemes. As future goal, we aim to extend the developed framework in order to analytically compute the case in which iterative interference cancellation is applied to received packets. Nonetheless, the use of asynchronous Aloha with diversity is of great interest in a wide range of application scenarios as for example Wireless Sensor Networks, in which first of all Interference Cancellation can not be applied due to battery constraints of the involved entities and secondly, synchronizing all terminals would require an unecessary payload and unecessary energy-consuming transmissions when terminals are waking up from sleeping mode.


\appendix

\textbf{Appendix}

In this section, the validity of the interference model is proven by demonstrating that the sum of all contributions is equal to 1. For simplicity, the demonstration is cut in 4 parts, each corresponding to the probability that the first event was (A), (B), (C) or (D) respectively and considering the union of the events for the second DP's copy. Therefore
\begin{equation}\begin{split}
Pr\{(A,CD)_{\tau'}\}+\sum_{x'=1}^{\tau'}Pr\{(B,BCD)_{x'}\}+\\ \\ +\sum_{x'=0}^{\tau'}Pr\{(C,ABCD)_{x'\}}+\sum_{x'=0}^{\tau'}Pr\{(D,ABCD)_{x'}\}=1
\end{split}\end{equation}\\
Any of the 4 terms are thoroughly discussed in the followings.\\ 
\begin{enumerate}[(i)]
\item \begin{equation}Pr\{(A,CD)_{\tau'}\}=\frac{1}{T_F'-\tau'+1}\end{equation}\\
represents the case in which the first copy of the DP entirely interfered with the CC. In this case, the two possible events for the second DP's copy are (C) and (D). This quantity has already been calculated and explained in Equation (13).\\

\item \begin{equation}\begin{split}
\sum_{x'=1}^{\tau'}Pr\{(B,BCD)_{x'}\}=Pr\{(B,B)_{\tau'}\}+\\ \\
+\sum_{x'=1}^{\tau'-1}Pr\{(B,CD)_{x'}\}+\sum_{x'=1}^{\tau'-1}Pr\{(B,B)_{x'}\}
\end{split}\end{equation}\\
represents the case in which the first copy of the DP partially interfered with the CC. In this case, the possible events for the second DP's copy are three: (B), (C) and (D). 
Substituting the values from Equations (17), (19), (20) and considering the common factors

\begin{equation}\begin{split}
\sum_{x'=1}^{\tau'}Pr\{(B,BCD)_{x'}\}=\\ \\
=\Bigg[\frac{1}{T_F'-\tau'+1}\cdot \frac{2}{T_F'-\tau'+1-(2\tau'-1)}\Bigg]\cdot\\ \\
\cdot\Bigg[(\tau'-1) + \sum_{x'=1}^{\tau'-1} \big[T_F'-\tau'+1-(2\tau'-1)-(\tau'-x')\big]+\\ \\
+\sum_{x'=1}^{\tau'-1} (x'-1)\Bigg]
\end{split}\end{equation}\\
that can be rewritten as\\
\begin{equation}\begin{split}
\sum_{x'=1}^{\tau'}Pr\{(B,BCD)_{x'}\}=\\ \\
=\Bigg[\frac{1}{T_F'-\tau'+1}\cdot \frac{2}{T_F'-\tau'+1-(2\tau'-1)}\Bigg]\cdot\\ \\
\Bigg[(\tau'-1) + (\tau'-1)\cdot \big[T_F'-\tau'+1-(2\tau'-1)\big]-\\ \\
-\sum_{x'=1}^{\tau'-1}(\tau'-x') + \sum_{x'=1}^{\tau'-1} (x'-1)\Bigg]
\end{split}\end{equation}\\

Considering that\\

\begin{equation}
-\sum_{x'=1}^{\tau'-1}(\tau'-x')=-(\tau'-1)-\sum_{x'=1}^{\tau'-1} (x'-1)
\end{equation}\\

Equation (36) becomes\\

\begin{equation}\begin{split}
\sum_{x'=1}^{\tau'}Pr\{(B,BCD)_{x'}\}=\\ \\
=\Bigg[\frac{1}{T_F'-\tau'+1}\cdot \frac{2}{T_F'-\tau'+1-(2\tau'-1)}\Bigg]\cdot\\ \\
\Bigg[(\tau'-1)\cdot [T_F'-\tau'+1-(2\tau'-1)]\Bigg]
\end{split}\end{equation}\\

Therefore
\begin{equation}\begin{split}
\sum_{x'=1}^{\tau'}Pr\{(B,BCD)_{x'}\}=\frac{2(\tau'-1)}{T_F'-\tau'+1}
\end{split}\end{equation}\\

\item 
\begin{equation}\begin{split}
\sum_{x'=0}^{\tau'}Pr\{(C,ABCD)_{x'\}}=Pr\{(C,A)_{\tau'}\}+\\ \\
\sum_{x'=1}^{\tau'-1}Pr\{(C,B)_{x'}\}+Pr\{(C,CD)_{0}\}
\end{split}\end{equation}

represents the case in which the first copy of the DP did not interfere with the CC and did not change the number of possible interference outcomes. In this case, the possible events for the second DP's copy are still four: (A), (B), (C) and (D).
Substituting the values from Equations (16), (21), (24) and considering the common factors

\begin{equation}\begin{split}
\sum_{x'=0}^{\tau'}Pr\{(C,ABCD)_{x'\}}=\Bigg[\frac{T_F'-\tau'+1-(4\tau'-1)}{T_F'-\tau'+1}\Bigg]\cdot \\ \\
\cdot \Bigg[\frac{1+\sum_{x'=1}^{\tau'-1}2+[T_F'-\tau'+1-2(2\tau'-1)]}{T_F'-\tau'+1-(2\tau'-1)}\Bigg]
\end{split}\end{equation}\\

With some simple mathematical passages it can be seen that the term on the second big square brackets equals $1$, therefore 

\begin{equation}
\sum_{x'=0}^{\tau'}Pr\{(C,ABCD)_{x'\}}=\frac{T_F'-\tau'+1-(4\tau'-1)}{T_F'-\tau'+1}
\end{equation}\\

\item 
\begin{equation}\begin{split}
\sum_{x'=0}^{\tau'}Pr\{(D,ABCD)_{x'}\}=Pr\{(D,A)_{\tau'}\} +\\ \\
+ \sum_{x'=1}^{\tau'-1}Pr\{(D,B)_x'\} + Pr\{(D,CD)_0\}
\end{split}\end{equation}\\

represents the case in which the first copy of the DP did not interfere with the CC but the number of possible interference outcomes is modified. In this case, the possible events for the second DP's copy are (A), (B), (C) and (D).
Substituting the values gathered from Equations (16), (22), (25) and considering the common factors

\begin{equation}\begin{split}
\sum_{x'=0}^{\tau'}Pr\{(D,ABCD)_{x'}\}=\\ \\
=\Bigg[\frac{1}{[T_F'-\tau'+1]\cdot [T_F'-\tau'+1-(2\tau'-1)]}\Bigg]\cdot\\ \\
\cdot \Bigg[2\tau' + \sum_{x'=1}^{\tau'-1} 2x' +\\ \\
+ \sum_{x'=1}^{\tau'-1} 4(\tau'-x') + \sum_{x'=1}^{\tau'-1} 2[T_F'-\tau'+1-(3\tau'+x'-1)]\Bigg]
\end{split}\end{equation}\\

Considering that\\ 

\begin{equation}
\sum_{x'=1}^{\tau'-1} x' = \sum_{x'=1}^{\tau'-1} (\tau'-x') = \frac{(\tau')\cdot(\tau'-1)}{2}
\end{equation}\\

equation (41) can be rewritten as\\

\begin{equation}
\begin{split}
\sum_{x'=0}^{\tau'}Pr\{(D,ABCD)_{x'}\}=\\ \\
=\Bigg[\frac{1}{[T_F'-\tau'+1]\cdot [T_F'-\tau'+1-(2\tau'-1)]}\Bigg]\cdot\\ \\
\Bigg[2\tau' + 2\frac{(\tau')\cdot(\tau'-1)}{2} + 4\frac{(\tau')\cdot(\tau'-1)}{2} +\\ \\
+ 2\tau'\cdot [T_F'-\tau'+1-(3\tau'-1)] - 2\frac{(\tau')\cdot(\tau'-1)}{2}\Bigg]
\end{split}
\end{equation}\\

from which\\

\begin{equation}\begin{split}
\sum_{x'=0}^{\tau'}Pr\{(D,ABCD)_{x'}\}=\\ \\
=\Bigg[\frac{1}{[T_F'-\tau'+1]\cdot [T_F'-\tau'+1-(2\tau'-1)]}\Bigg]\cdot\\ \\
\Bigg[2\tau' \cdot [T_F'-\tau'+1-(2\tau'-1)]\Bigg]
\end{split}\end{equation}\\

and simplifying

\begin{equation}\begin{split}
\sum_{x'=0}^{\tau'}Pr\{(D,ABCD)_{x'}\}=\frac{2\tau'}{T_F'-\tau'+1}
\end{split}\end{equation}\\

\end{enumerate}

It can be seen that the sum of the terms found in Equations (30),(36),(39) and (45) is equal to $1$. This validates Equation (29).

\end{document}